\documentclass[12pt]{article}
\usepackage{a4wide}
\usepackage{graphicx}

\newcommand{\order}{\ensuremath{{\cal O}}}

\newcommand{\be}{\begin{equation}}
\newcommand{\ee}{\end{equation}}
\newcommand{\ba}{\begin{eqnarray}}
\newcommand{\ea}{\end{eqnarray}}

\begin{document}
\begin{titlepage}
\begin{flushright}
LU TP 03-10\\
hep-ph/0303103\\
revised June 2003
\end{flushright}
\vfill
\begin{center}
{\Large \bf \boldmath{$K_{\ell3}$} decays in Chiral Perturbation Theory}
\vfill
{\bf Johan Bijnens}\\[0.3cm]
{Department of Theoretical Physics 2, Lund University,\\
S\"olvegatan 14A, S 223-62 Lund, Sweden}\\[1cm]
{\bf Pere Talavera}\\[0.3cm]
Departament de F{\'\i}sica i Enginyeria Nuclear,\\
Universitat Polit\`ecnica de Catalunya,\\
Jordi Girona 1-3, E-08034 Barcelona, Spain
\end{center}

\vfill

\begin{abstract}
The process $K_{\ell3}$ is calculated to two-loop order ($p^6$)
in Chiral Perturbation Theory (ChPT)
in the isospin conserved case.
We present expressions suitable for use with previous work
in two-loop CHPT where the order $p^4$ parameters ($L_i^r$) were
determined from experiment. We point out that all the order $p^6$ parameters
($C_i^r$) that appear in the value of $f_+(0)$ relevant for the
determination
of $|V_{us}|$ can be determined from $K_{\ell3}$ measurements
via the slope and the curvature of the scalar form-factor.
As by product we update the value of the CKM matrix element
$\vert V_{us} \vert$.
\end{abstract}
\vfill
{\bf PACS:} 12.15.Hh,13.20.Eb,12.39.F,14.40.Aq
\vfill
\footnoterule
{\footnotesize\noindent$^\dagger$ Supported in part by the European
Union TMR
network, Contract No. HPRN-CT-2002-00311  (EURIDICE)}

\end{titlepage}

\section{Introduction}

Weak semileptonic kaon decays to a pion and a
lepton-neutrino pair ($K_{\ell3}$)
have a long history. The early theoretical treatments can be found in the
review \cite{kl3old}. This decay plays an important role in the
determination of the CKM matrix-element $V_{us}$, see e.g. the discussion
in \cite{PDG2002} or \cite{Calderon}.
The theoretical basis for this determination is
the paper by Leutwyler and Roos \cite{LR}. The basis for this
evaluation were radiative corrections and estimates of one-loop
chiral corrections. The full Chiral Perturbation Theory calculation
to order $p^4$ (see Sect.~\ref{definitions} for a short explanation)
was performed by Gasser and Leutwyler~\cite{GL2}.
References
to earlier work on the non-analytic corrections can be found there.
Recent Reviews of the situation can be found in \cite{daphne} or
\cite{kaon99}.

Since that time a lot of work has been performed in Chiral Perturbation
Theory. An update of the calculation of \cite{GL2} to order $p^6$ is thus
necessary. Partial studies are done, the double logarithm contribution
is small \cite{BCE2} and a possibly large role for terms with two
powers of quark masses has been argued for in Ref.~\cite{FKS}.
One full order $p^6$ calculation exists \cite{Post3}, but
it uses outdated values of the ChPT constants as well as an older version
of the classification of $p^6$ constants.
In this paper we
present an independent calculation of the $K_{\ell3}$ amplitudes
to order $p^6$ in ChPT in the isospin limit. We present numerical
results with values for the ChPT constants resulting from fits to
order $p^6$~\cite{ABT1,ABT2,ABT3,BT1}. Related work is the update
of the electromagnetic radiative corrections given in \cite{Radiative}.

We present a few definitions of ChPT in Sect.~\ref{definitions}
to determine our notation. Sect.~\ref{defKl3} defines the form-factors
used in $K_{\ell3}$ decays.
Analytical results are presented explicitly in Section~\ref{analytical}
for the form-factors up to order $p^4$ and for the
part depending on the order $p^6$ parameters ($C_i^r$). The remaining parts
are rather long and can be obtained from the authors on request.
Section~\ref{fp0Ci} describes one of our main results, the fact that all
needed $p^6$ constants for the value of $f_+(0)$ can be experimentally
determined from $K_{\mu3}$ experiments.
The value of $f_+(0)$ is of course needed for determinations of
$V_{us}$ and is of use for future precise measurements of
$K\to\pi\nu\bar\nu$.

In Section~\ref{data} the presently available data set is discussed.
Here we also point out that the often used linear approximation in the
form-factors can have a sizable effect on the measured value
of the slope and the value at $t=0$.
This effect is of similar size as the experimental errors quoted.
We present an extended discussion of the numerical results in
Sect.~\ref{numerics} after a short discussion of the inputs used in
Sect.~\ref{inputs}. Our final conclusions for $\lambda_+$ are in
Sect.~\ref{valuel+} and of $f_+(0)$ in Sect.~\ref{valuefp0}.
We summarize our results in Section~\ref{conclusions}.

\section{Some definitions}
\label{definitions}
\setcounter{equation}{0}

\begin{figure}
\begin{center}
\includegraphics[width=0.6\textwidth]{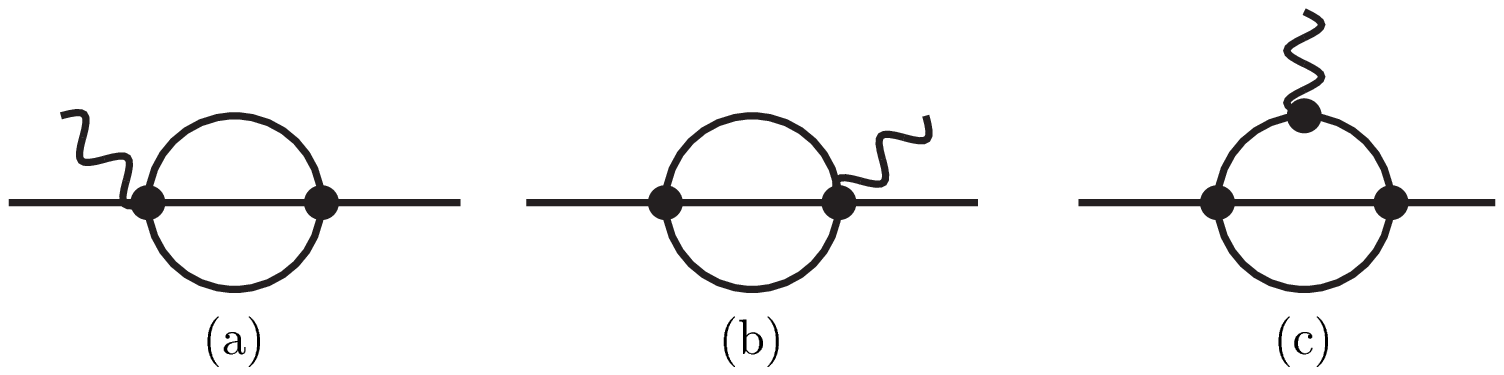}
\end{center}
\caption{\label{figtwoloop} The two-loop diagrams
with overlapping loop integrals contributing to the
$K_{\ell3}$ form-factor. A $\bullet$ indicates a vertex from ${\cal L}_2$,
a {\boldmath$\times$} a vertex from  ${\cal L}_4$ and
a {\boldmath$\otimes$} a vertex from ${\cal L}_6$. The insertion of the
weak current is indicated by the wiggly line.}
\end{figure}

Chiral Perturbation Theory
is the modern way to derive the predictions
following from the fact the $SU(n_f)_L\times SU(n_f)_R$
chiral symmetry in the limit of $n_f$ massless flavours in QCD
is  spontaneously
broken by nonperturbative QCD dynamics to the diagonal vector subgroup
$SU(n_f)_V$.
It is the effective field theory method
to use this property at low energies.
It takes into account the
singularities associated with the Goldstone Boson degrees of freedom
caused by the spontaneous breakdown of chiral symmetry and
parametrizes all the remaining freedom allowed by the chiral Ward
identities in low energy constants (LECs). The LECs are the freedom in
the parts of the amplitudes that depend analytically on the masses and
momenta.  The expansion is ordered in terms of momenta, quark
masses and external fields. Recent lectures introducing this area are
given in ref.~\cite{chptlectures}.
We use here the standard ChPT counting where the
quark mass, scalar and pseudoscalar external fields are counted as two
powers of momenta. Vector and axial-vector external currents count as
one power of momentum. The lagrangian can be ordered as
\begin{eqnarray}
\label{EffLag}
{\cal L}^{\rm effective} &=&  {\cal L}_2 + {\cal L}_4 + {\cal L}_6 +\cdots
\nonumber\\
&=& {\cal L}_2 + \sum_{i=1}^{10} L_i \; O_4^i
 + \sum_{i=1}^{90} C_i \; O_6^i + \sum_{i=91}^{94} C_i \;
 + \cdots\,.
\end{eqnarray}
The index $i$ in ${\cal L}_i$ stands for the chiral power.  The
precise form of ${\cal L}_2$ and ${\cal L}_4$ is given below
while
${\cal L}_6$ can be found
in~\cite{BCE1}.
The lowest order, $\order(p^2)$, in the expansion
corresponds to tree level diagrams with vertices from ${\cal L}_2$,
the next-to-leading order, NLO or $\order(p^4)$, to one-loop diagrams
with vertices from ${\cal L}_2$ or tree level diagrams with one vertex
from ${\cal L}_4$ and the rest from ${\cal L}_2$. The
next-to-next-to-leading order, NNLO or $\order(p^6)$, has two-loop
diagrams, one-loop diagrams with one vertex from ${\cal L}_4$ and tree
level diagrams with one vertex from ${\cal L}_6$ or two vertices from
${\cal L}_4$ and all other vertices from ${\cal L}_2$.  The loop
diagrams take all singularities due to the Goldstone Bosons correctly
into account.  The singularities are the real predictions of ChPT
while the other effects from QCD are in the values of the LECs.
The diagrams, in addition to wave-function-renormalization, relevant
for the processes discussed in this paper are shown in
Figs.~\ref{figtree}, \ref{figoneloop} and \ref{figtwoloop}.

The expressions for the first two terms in the expansion of the
Lagrangian are given
by ($F_0$ refers to the
pseudoscalar decay constant in the chiral limit)
\begin{equation}
\label{L2}
{\cal L}_2 = \frac{F_0^2}{4} \{\langle D_\mu U^\dagger D^\mu U \rangle
+\langle \chi^\dagger U+\chi U^\dagger \rangle \}\, ,
\end{equation}
and
\begin{eqnarray}
\label{L4}
{\cal L}_4&& =
L_1 \langle D_\mu U^\dagger D^\mu U \rangle^2
+L_2 \langle D_\mu U^\dagger D_\nu U \rangle
     \langle D^\mu U^\dagger D^\nu U \rangle \nonumber\\&&\hspace{-0.5cm}
+L_3 \langle D^\mu U^\dagger D_\mu U D^\nu U^\dagger D_\nu U\rangle
+L_4 \langle D^\mu U^\dagger D_\mu U \rangle \langle \chi^\dagger U
+\chi U^\dagger \rangle
\nonumber\\&&\hspace{-0.5cm}
+L_5 \langle D^\mu U^\dagger D_\mu U (\chi^\dagger U+U^\dagger \chi )
\rangle
+L_6 \langle \chi^\dagger U+\chi U^\dagger \rangle^2
\nonumber\\&&\hspace{-0.5cm}
+L_7 \langle \chi^\dagger U-\chi U^\dagger \rangle^2
+L_8 \langle \chi^\dagger U \chi^\dagger U
+ \chi U^\dagger \chi U^\dagger \rangle
\nonumber\\&&\hspace{-0.5cm}
-i L_9 \langle F^R_{\mu\nu} D^\mu U D^\nu U^\dagger +
               F^L_{\mu\nu} D^\mu U^\dagger D^\nu U \rangle
+L_{10} \langle U^\dagger  F^R_{\mu\nu} U F^{L\mu\nu} \rangle
\,,
\end{eqnarray}
while the next-to-next-to-leading order is a rather cumbersome expression
\cite{BCE1}.
The special unitary matrix $U$ contains the Goldstone boson fields
\be
U = \exp\left(\frac{i\sqrt{2}}{F_0}M\right)\,,\quad
M =\left(\begin{array}{ccc}
\frac{1}{\sqrt{2}}\pi^0+\frac{1}{\sqrt{6}}\eta & \pi^+ & K^+\\
\pi^- & \frac{-1}{\sqrt{2}}\pi^0+\frac{1}{\sqrt{6}}\eta & K^0\\
K^- & \overline{K^0} & \frac{-2}{\sqrt{6}}\eta
         \end{array}\right)\,.
\ee
The formalism is the external field method of \cite{GL1}
with $s$, $p$, $l_\mu$ and $r_\mu$ matrix valued scalar, pseudo-scalar,
left-handed and right-handed vector external fields respectively.
These show up in
\be
\chi = 2 B_0\left(s+ip\right)\,,
\ee
in the covariant derivative
\be
\label{covariant}
D_\mu U = \partial_\mu U -i r_\mu U + i U l_\mu\,,
\ee
and in the field strength tensor
\be
F_{\mu\nu}^{L(R)} = \partial_\mu l(r)_\nu -\partial_\nu l(r)_\mu -i
\left[ l(r)_\mu , l(r)_\nu \right]\,.
\ee
For our purpose it is sufficient to set
\be
s =
\left(\begin{array}{ccc}\hat m &  & \\ & \hat m & \\ & &
m_s\end{array}\right)\,,
\quad
r_\mu = 0\,,\quad l_\mu =
-\frac{g_2}{\sqrt2}
\left(\begin{array}{ccc}
 & V_{ud}W^+_\mu  & V_{us}W^+_\mu\\
 V_{ud}^* W^-_\mu & & \\
 V_{us}^* W^-_\mu& &
\end{array}\right)
\ee
with $g_2$ the weak coupling constant, related to the Fermi constant by
\be
\frac{G_F}{\sqrt{2}} = \frac{g_2^2}{8 m_W^2}\,.
\ee

\subsection{Renormalization Scheme}

We use the renormalization scheme as explained in \cite{BCE1} and
\cite{BCEGS}. It extends the scheme from \cite{GL1} naturally to two-loops.
Notice that the work of Post and Schilcher \cite{Post3,Post1,Post2} used
a slightly different scheme. The scheme employed here
does not introduce the $\epsilon^2$ term in Eq.~(39) of Ref.~\cite{Post3}.
Subtractions are performed via
\ba
L_i &=& (C\mu)^{(d-4)}\left(\Gamma_i+L_i(\mu)\right)\,,
\nonumber\\
C_i &=& (C\mu)^{2(d-4)}\left(C_i^r(\mu)-\frac{1}{F^2}
\left[\Gamma_i^{(2)}\Lambda^2+( \Gamma_i^{(1)}+\Gamma_i^{(L)}(\mu))\Lambda
\right]\right)\,,
\ea
with
\be
\ln C = -\frac{1}{2}\left(\ln(4\pi)+\Gamma^\prime(1)+1\right),\quad
\Lambda = \frac{1}{16\pi^2(d-4)}\,.
\ee
The coefficients $\Gamma_i$ can be found in \cite{GL1,BCE1} and the
$\Gamma_i^{(2)}, \Gamma_i^{(1)}\Gamma_i^{(L)}(\mu)$ in \cite{BCE1}.
We will in the remainder always write $C_i^r$ and $L_i^r$ but the
$\mu$ dependence is of course present.

\section{The \boldmath$K_{\ell3}$ form-factors: definition and
\boldmath$\order(p^4)$·}
\label{defKl3}
\setcounter{equation}{0}

The decays we consider are
\ba
 K^+(p) &\rightarrow& \pi^0 (p') \ell^+ (p_\ell) \nu _\ell (p_\nu)\,,
 \hspace{1cm}
[K_{\ell3}^+] \label{decayKp}
\\
K^0(p) &\rightarrow &\pi^- (p') \ell^+ (p_\ell) \nu_\ell (p_\nu)\,,
\hspace{1cm}
[K_{\ell3}^0]
\label{decayKo}
\ea
and their charge conjugate modes.
$\ell$ stands for $\mu$ or $e$.

The matrix-element for $K_{\ell3}^+$,
neglecting scalar and tensor contributions,
has the structure
\ba
T& =& \frac{G_F} {\sqrt{2}} V_{us}^\star \ell^\mu { F_\mu}^+ (p',p)\,,
\label{s34}
\ea
with
\ba
\ell^\mu& =& \bar{u} (p_\nu)\gamma^\mu  (1- \gamma_5) v (p_\ell)\,,
\nonumber \\
{ F_\mu}^+ (p',p)& =& < \pi^0 (p') \mid V_\mu^{4-i5} (0)
\mid K^+(p)>\,,
\nonumber \\
&=& \frac{1}{\sqrt{2}} [(p'+p)_\mu f^{K^+\pi^0}_+ (t) + (p-p')_\mu
f_-^{K^+\pi^0} (t)]. \label{s35}
\ea
To obtain the $K_{\ell3}^0$ matrix-element, one replaces $F^+_\mu$ by
\ba
{F_\mu }^0 (p',p)& =& < \pi^- (p') \mid V_\mu^{4-i5} (0)
\mid K^0(p)>
\nonumber \\
&=&  (p'+p)_\mu f^{K^0\pi^-}_+ (t) + (p-p')_\mu f_-^{K^0\pi^-} (t).
\label{s35a}
\ea
 The processes (\ref{decayKp})
and (\ref{decayKo}) thus involve the four $K_{\ell3}$ form-factors
$f^{K^+\pi^0}_\pm (t)$, $f^{K^0 \pi^-}_\pm (t)$ which depend on
\be
t = (p'-p)^2 = (p_\ell + p_\nu)^2,
\label{s36}
\ee
the square of the four momentum transfer to the leptons.

In this paper we work in the isospin limit thus
\be
f_\pm = f_\pm^{K\pi}=f_\pm^{K^+\pi^0} = f_\pm^{K^0\pi^-}\,.
\ee
$f_+^{K\pi}$ is referred to as the vector form-factor, because
it specifies the $P$-wave  projection of the crossed channel matrix-elements
 $< 0 \mid V^{4-i5}_\mu(0) \mid K^+, \pi^0 \;\mbox{in} >$.
 The $S$-wave projection is described by the scalar form-factor
\be
f_0 (t) = f_+ (t) + \frac{t}{m^2_K - m^2_\pi} f_-(t)
\,.
 \label{s37}
\ee

Analyses of $K_{\ell3}$ data frequently assume a linear dependence
\be
f_{+,0} (t) = f_+ (0) \left[ 1 + \lambda_{+,0}
\frac{t}{m^2_{\pi^+}} \right] \; \; .
\label{s38}
\ee
For a discussion of the validity of this approximation see \cite{GL2}
and references cited therein.
We will discuss it to order $p^6$ and in comparison with the data.
At the expected future precision it will be necessary to go
beyond this approximation.

Eqs. (\ref{s37}) and (\ref{s38}) leads to a constant
$f_- (t)$ ,
\be
f_- (t) = f_- (0) = f_+ (0) (\lambda_0 -
\lambda_+) \frac{m_K^2 - m_\pi^2}{m_{\pi^+}^2}.
\label{s381}
\ee
The form-factors $f_{\pm,0} (t)$ are analytic functions in the complex
$t$-plane cut along the positive real axis. The cut starts at $t=(m_K +
m_\pi)^2$. In our phase convention, the form-factors are real in the
physical
region
\be
m^2_\ell \leq t \leq (m_K - m_\pi)^2.
\label{s39}
\ee
A discussion of the kinematics in $K_{\ell3}$ decays can be found
in \cite{daphne}.

\section{Analytical Results}
\label{analytical}
\setcounter{equation}{0}

The total result we obtain is split by chiral order.
\be
f_i(t) = f_i^{(2)}(t)+f_i^{(4)}(t)
+f_i^{(6)}(t)\,, \quad (i=+,-,0)\,.
\ee

\subsection{Order \boldmath $p^2$}

\begin{figure}
\begin{center}
\includegraphics[width=0.6\textwidth]{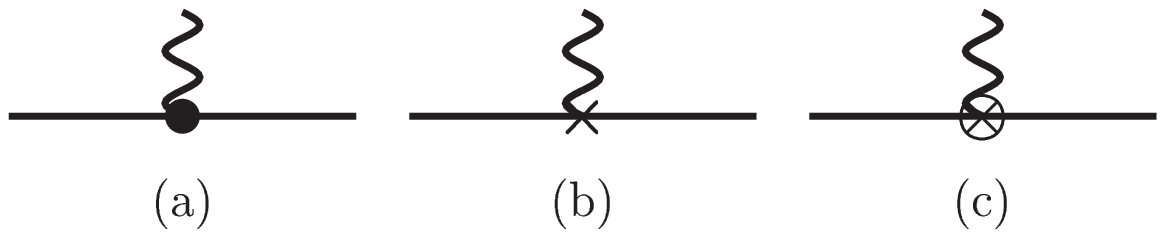}
\end{center}
\caption{\label{figtree} The tree level diagrams contributing to the
$K_{\ell3}$ form-factor. A $\bullet$ indicates a vertex from ${\cal L}_2$,
a {\boldmath$\times$} a vertex from  ${\cal L}_4$ and
a {\boldmath$\otimes$} a vertex from ${\cal L}_6$. The insertion of the
weak current is indicated by the wiggly line.}
\end{figure}

This has been known for a very long time and is fully
determined by gauge invariance.
\be
f_+^{(2)}(t) = f_0^{(2)}(t) = 1,,\quad\quad\quad
f_-^{(2)}(t) = 0\,.
\ee
It results from
the diagram in Fig.~\ref{figtree}(a).

\subsection{Order \boldmath $p^4$}

\begin{figure}
\begin{center}
\includegraphics[width=0.7\textwidth]{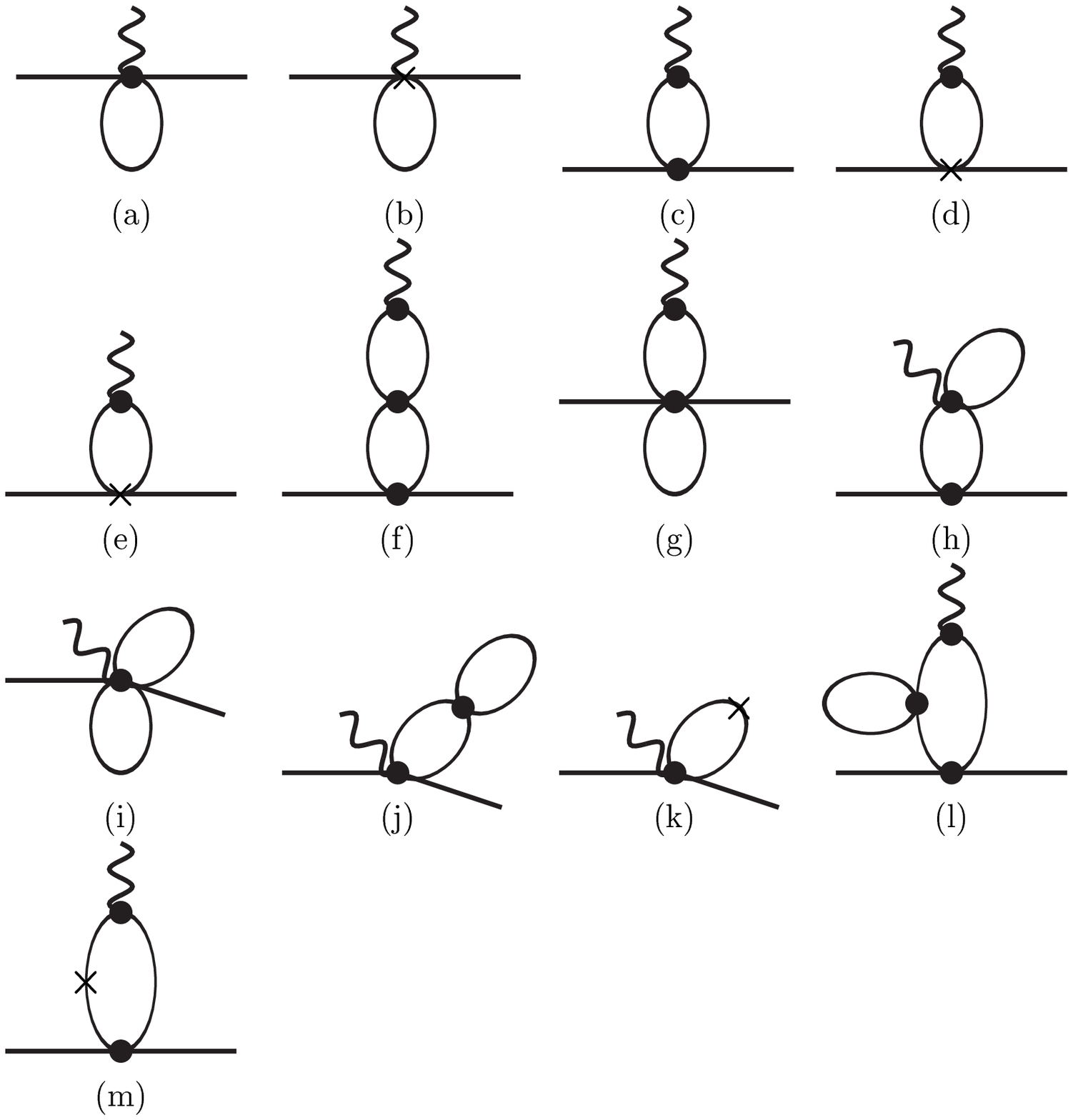}
\end{center}
\caption{\label{figoneloop} The one and two-loop diagrams
without overlapping loop integrals contributing to the
$K_{\ell3}$ form-factor. A $\bullet$ indicates a vertex from ${\cal L}_2$,
a {\boldmath$\times$} a vertex from  ${\cal L}_4$ and
a {\boldmath$\otimes$} a vertex from ${\cal L}_6$. The insertion of the
weak current is indicated by the wiggly line.}
\end{figure}

This was first calculated within the ChPT framework
by Gasser and Leutwyler in 1985 \cite{GL2}.
The result contains the nonanalytic dependence in the symmetry parameters
predicted by \cite{DashenLi}.
The form of the form-factors
we use (which is equivalent to the result of \cite{GL2}
to order $p^4$)
is the one which our expressions for the $p^6$ contribution
correspond to.

\ba
F_\pi^2\,f_+^{(4)}(t) &=&
   2L_9^r t
   + 3/8\left(
\overline{A}(m_\eta^2) + \overline{A}(m_{\pi}^2) +
2\overline{A}(m_K^2)\right)
\nonumber\\&&
       -3/2\,\left( \overline{B}_{22}(m_{\pi}^2,m_K^2,t)
                  + \overline{B}_{22}(m_K^2,m_\eta^2,t)\right)\,.
\ea

\ba
F_\pi^2\,f_-^{(4)}(t) &=&
 (4 m_K^2 - 4 m_{\pi}^2) L_5^r - 2  (m_K^2 - m_{\pi}^2) L_9^r
   + 1/2 \overline{A}(m_\eta^2)
   - 5/12 \overline{A}(m_{\pi}^2)
\nonumber\\&&
   + 7/12 \overline{A}(m_K^2)
   + \overline{B}(m_{\pi}^2,m_K^2,t)(  - 1/12 m_{\pi}^2 - 5/12 m_K^2 +
5/12 t)
\nonumber\\&&
   + \overline{B}(m_K^2,m_\eta^2,t) ( 1/12 m_{\pi}^2 - 7/12 m_K^2 + 1/4 t )
\nonumber\\&&
  + \overline{B}_1(m_{\pi}^2,m_K^2,t)(- 7/12 m_{\pi}^2 + 19/12 m_K^2 -
5/12 t )
\nonumber\\&&
  + \overline{B}_1(m_K^2,m_\eta^2,t)(- 11/12 m_{\pi}^2 + 23/12 m_K^2 -
1/4  t )
\nonumber\\&&
  + \overline{B}_{21}(m_{\pi}^2,m_K^2,t) ( 3/2 m_{\pi}^2 - 3/2 m_K^2 -
5/6 t )
\nonumber\\&&
  + \overline{B}_{21}(m_K^2,m_\eta^2,t)  ( 3/2 m_{\pi}^2 - 3/2 m_K^2 -
1/2 t )
\nonumber\\&&
  + \overline{B}_{22}(m_{\pi}^2,m_K^2,t)    (  - 5/6 )
  + \overline{B}_{22}(m_K^2,m_\eta^2,t)    (  - 1/2 )\,.
\ea
These results are obtained from the diagrams in Figs.~\ref{figtree}(b),
\ref{figoneloop}(a) and \ref{figoneloop}(c), together with wave-function
renormalization.

\subsection{Order $p^6$}

The $p^6$ contribution we split in several parts
\be
f_i^{(6)}(t) = \frac{1}{F_\pi^4}
\left(f_i^C(t)+f_i^L(t)+f_i^B(t)+f_i^H(t)+f_i^V(t)\right)\,, \quad
(i=+,-,0)\,.
\ee
The split between the last three terms is not unique and depends on
how the irreducible two-loop integrals are separated from the reducible
ones. The first two terms are the ones containing the dependence
on the $p^6$ and $p^4$ coupling constants.

The ones with dependence on $C_i^r$ stem from wave-function
renormalization and
the diagram of Fig.~\ref{figtree}(c).
The results are
\ba
\label{fpCi}
f_+^C(t) &\equiv& R_{+0}^{K\pi} + t R_{+1}^{K \pi}
 +t^2 R_{+2}^{K\pi}
\nonumber\\
&=& -8 \left(m_\pi^2-m_K^2\right)^2\,\left(C_{12}^r+C_{34}^r\right)
+ t \Big[ -4 m_\pi^2 (  2 C_{12}^r +4 C_{13}^r + C_{64}^r
\nonumber\\&&
                 + C_{65}^r + C_{90}^r )
     +m_K^2( - 8 C_{12}^r - 32  C_{13}^r - 8  C_{63}^r - 8  C_{64}^r
 - 4  C_{90}^r )\Big]
\nonumber\\&&
+ t^2  (  - 4 C_{88}^r + 4 C_{90}^r )\,,
\ea
and
\ba
f_-^C(t) &\equiv& R_{-0}^{K\pi} + t R_{-1}^{K \pi}
 +t^2 R_{-2}^{K\pi}
\nonumber\\
&=&   \left(m_K^2-m_\pi^2\right)
  \Big[m_\pi^2(+24C_{12}^r -16C_{13}^r +8C_{15}^r +16C_{17}^r +8C_{34}^r
  +4C_{64}^r
\nonumber\\&&
+4C_{65}^r+4C_{90}^r )
         +m_K^2  (+24C_{12}^r+32C_{13}^r +16C_{14}^r+16C_{15}^r +8C_{34}^r
\nonumber\\&&
 +4C_{63}^r +8C_{64}^r +4C_{90}^r )\Big]
       + t
4\left(m_K^2-m_\pi^2\right)\left(-2C_{12}^r+C_{88}^r-C_{90}^r\right)
\,.
\ea

We have followed a notation very close to the one in \cite{BT1}.
Notice that we have the relations
\ba
R_{-2}^{K\pi} &=&0\,,
\nonumber\\
R_{+2}^{K\pi} &=& R_{V2}^{\pi} = R_{V2}^{K^+}\,,
\nonumber\\
R_{+1}^{K\pi} &=& \frac{1}{2}\left( R_{V1}^\pi + R_{V1}^{K^+}\right)
                 + R_{V1}^{K^0}\,.
\ea
The other $R_i^M$ are similar combinations of the $C_i^r$
but in the expansion of the
electromagnetic form-factors \cite{BT1}.
Notice that the last relation is really Sirlin's relation \cite{Sirlin}
and the second satisfies it as well.

We have not quoted the remaining formulas, but will quote below
some approximate numerical expressions. The exact formulas can be
obtained from the authors on request. Our expressions satisfy the
Ademollo-Gatto theorem~\cite{AG}.

\section{Getting the value of \boldmath$f_+(0)$}
\label{fp0Ci}
\setcounter{equation}{0}

One of the problems we face here is whether the needed $C_i^r$ can
be determined from experiment.
There are many of these coefficients showing up but as is obvious
from Eq.~(\ref{fpCi}), what we need is a value for $C_{12}^r+C_{34}^r$.
It turns out that this combination can actually be determined from
$K_{\ell3}$ measurements. The derivation given below relies on the
fact that we need values for the $p^4$ constants, determined to
order $p^4$ only, in the order $p^6$ part to be correct to the accuracy
that we
are working. We can determine all needed $L_i^r$ to this accuracy
relying only
on data.

We construct the quantity
\be
\label{deftildef0}
\tilde f_0(t) = f_+(t)+\frac{t}{m_K^2-m_\pi^2}
\left(f_-(t)+1-F_K/F_\pi\right)
= f_0(t)+\frac{t}{m_K^2-m_\pi^2}\left(1-F_K/F_\pi\right)
\,.
\ee
This has no dependence on the $L_i^r$ at order $p^4$, only via
order $p^6$ contributions. Inspection of the dependence on the $C_i^r$ shows
that
\ba
\label{resultfp0}
\tilde f_0(t) &=& 1-\frac{8}{F_\pi^4}\left(C_{12}^r+C_{34}^r\right)
\left(m_K^2-m_\pi^2\right)^2
+8\frac{t}{F_\pi^4}\left(2C_{12}^r+C_{34}^r\right)\left(m_K^2+m_\pi^2\right)
\nonumber\\&&
-\frac{8}{F_\pi^4} t^2 C_{12}^r
+\overline\Delta(t)+\Delta(0)\,.
\ea
We emphasize that the quantities $\overline\Delta(t)$ and
$\Delta(0)$ can in principle be calculated
to order $p^6$ accuracy with knowledge of the $L_i^r$ to order $p^4$
accuracy. In practice, since a $p^4$ fit will include in the values of the
$L_i^r$ effects that come from the $p^6$ loops (due to the fitting to
experimental values)
 we consider the $p^6$
fits to be the relevant ones to avoid double counting effects.
 Numerical results will be discussed in Section ~\ref{numerics}.

The definition in (\ref{deftildef0}) has essentially used the
Dashen-Weinstein
relation \cite{DW}
to remove the $L_i^r$ dependence at order $p^4$. It has also the
effect that it removed many of the $C_i^r$ from the scalar form-factor
as well. The corrections which appear in the Dashen-Weinstein relation
are include in the functions $\overline\Delta(t)$ and $\Delta(0)$,
these have both order $p^4$ \cite{GL2,DashenLi} and order $p^6$
contributions.

It is obvious from Eq.~(\ref{resultfp0}) that the needed combination
of $C_i^r$ can be determined from the slope and the
curvature of the scalar form-factor in $K_{\ell3}$ decays.

It seems possible that
$C_{12}^r$ can be measured from the curvature of the pion scalar form-factor
near 0 \cite{BD}. When this calculation is complete, one can use the
dispersive estimates of the pion scalar form-factor together with only a
$\lambda_0$ measurement in $K_{\mu3}$ to obtain the $p^6$ value for
$f_+(0)$. There are also some dispersive estimates for the relevant
scalar form-factor. Unfortunately, these are not in a usable form at
present \cite{Pich2}.

The discussion of Ref.~\cite{FKS} can be shown in this light too.
The constant $A_3$ of generalized perturbation theory
correspond to a combination of the $C_i^r$ from \cite{BCE1}.
The precise combination is
\be
A_3 = (2B_0)^2\left[- C_{34}^r+\frac{1}{2}
\left(C_{14}^r+C_{17}^r+C_{26}^r+C_{29}^r+C_{31}^r\right)+\frac{1}{6}C_{33}^r
\right]\,.
\ee
As we have shown it is possible to eliminate all but $C_{34}^r$, so this
plays
the role of $A_3$ here.
$C_{12}^r$ is higher order in the counting employed in \cite{FKS}
and was not considered there.

In Ref.~\cite{FKS} the additional observation was made
that a relatively small change in
the ratio $F_K/F_\pi$, together with an adjustement of the constant $A_3$
can accommodate the CKM unitarity.

\section{Data}
\label{data}
\setcounter{equation}{0}

\subsection{\boldmath $K^0_{\mu3}$}

The most useful data points are those where the full dependence
on the kinematical variable $t= q^2/m_\pi^2$ is shown.
That means that the experiments
that determined the value of $\lambda_0$ from the branching ratio
or did not provide an actual $t$ dependence but just fitted the linear slope
in a global way are not that useful for us, but see below.

Of the more recent experiments that quote data not from the branching ratio,
the ones that gave a plot or numbers for $f_0(t)$ are
\cite{Donaldson74B,Clark77,Birulev81,Buchanan75}. Ref.~\cite{Hill79}
gives a plot but mentions that it is not statistically significant, which
inspection of the plots confirms.

\begin{figure}
   \centering
   \includegraphics[angle=270,width=0.8\textwidth]{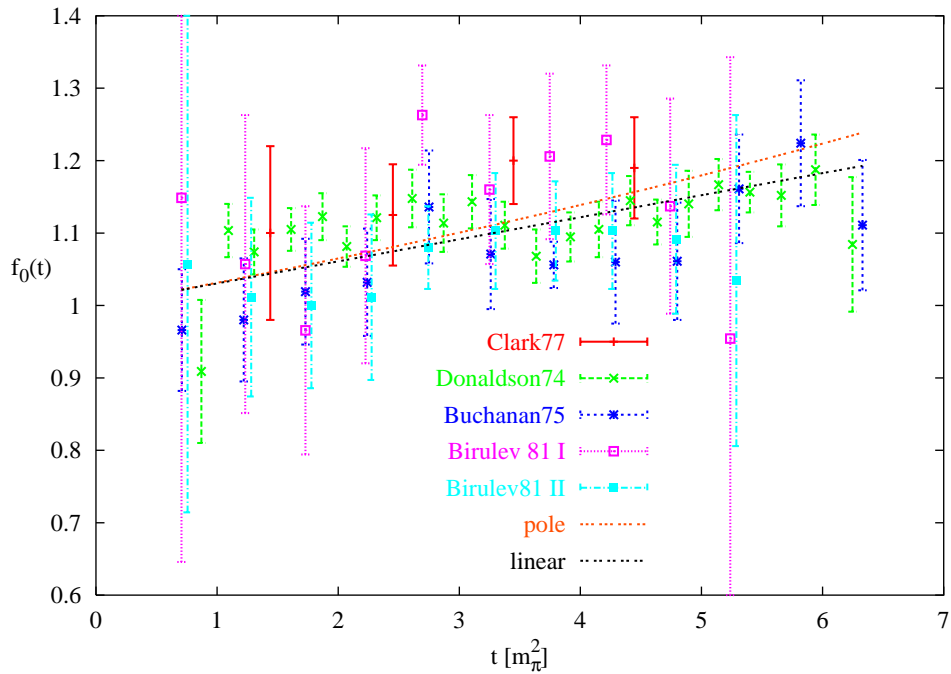}
   \caption{The data on $f_0(t)/f_+(0)$ from $K^0_{\mu3}$.
The data are Clark77 \cite{Clark77}, Donaldson74 \cite{Donaldson74B},
Buchanan75  \cite{Buchanan75} and Birulev81  \cite{Birulev81},
The latter reference has two distinct data sets. For comparison
a linear approximation and a pole approximation with a mass of 800~MeV are
shown as well.}
   \label{fig:f0data}
\end{figure}

There are some obvious problems with the data.
E.g. the $f_0(t)$ from \cite{Donaldson74B} do not go to 1 at $t=0$.
We have shown these data in Fig.~\ref{fig:f0data} together
with a linear and a pole approximation corresponding to
a mass of 800~MeV. This shows the accuracy needed to see the curvature.

\subsection{\boldmath $K^+_{\mu3}$}

Here we have not been able to find data that show a plot of $f_0(t)$.
All experiments are analyzed in terms of a constant form-factor as discussed
in \cite{PDG2002}. There is one more experiment \cite{Ajinenko03} that
quotes
measurements of $\lambda_0$ not included in \cite{PDG2002}.

\subsection{\boldmath $K^0_{e3}$}
\label{dataK0e3}

\begin{figure}
   \centering
   \includegraphics[angle=270,width=0.8\textwidth]{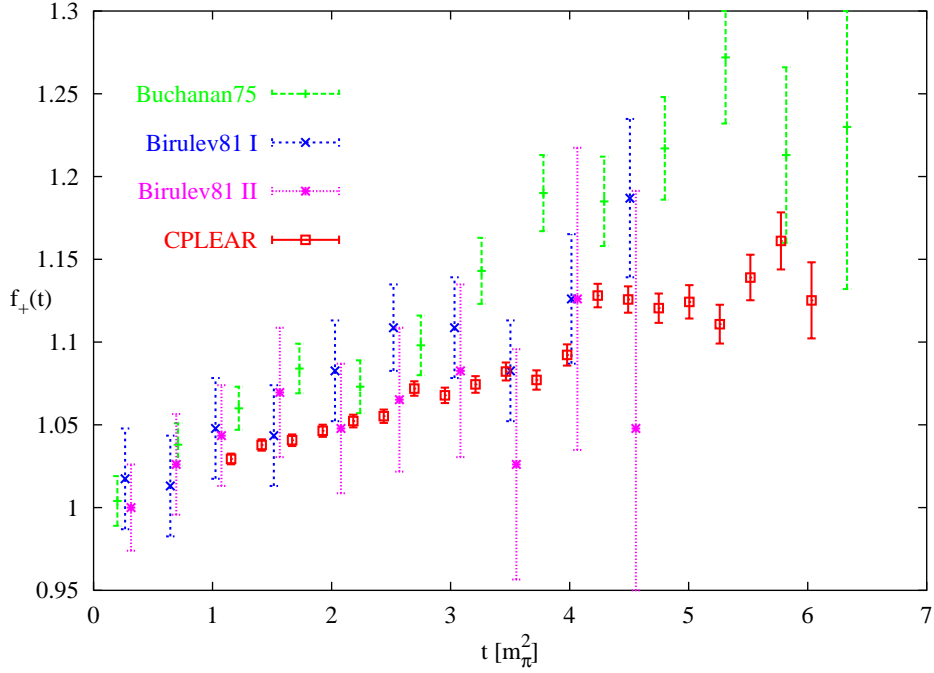}
   \caption{The data on $f_+(t)/f_+(0)$ from $K^0_{e3}$.
The CPLEAR data are from Ref.~\cite{Apostolakis00}. The older
data are for comparison, Buchanan75 is \cite{Buchanan75} and Birulev81
is \cite{Birulev81}. The latter reference has two distinct data sets.}
   \label{fig:fpdata}
 \end{figure}

Here the data are dominated by the recent high statistics CPLEAR data
\cite{Apostolakis00}. There exists a very high statistics older experiment
\cite{Gjesdal76}. They provide plots with different data assumptions
and can thus not be easily compared at the level of $f_+(t)$ directly.
But \cite{Gjesdal76} quoted both a linear and quadratic fit to $f_+(t)$.
In order to show the relation between the most recent and older data we have
plotted the data of \cite{Buchanan75}, \cite{Birulev81} and
\cite{Apostolakis00} in Fig.~\ref{fig:fpdata}.

We have performed some simple fits of the form
\be
\label{fitfpsimple}
f_+(t) = a_+ \left(1+ \lambda_+ \frac{t}{m_{\pi^+}^2} + c_+ t^2\right)
\ee
to the CPLEAR data. The fits agree extremely well with those reported
in \cite{Apostolakis00} and are given in Table~\ref{tab:fitsfp}.
\begin{table}
  \centering
  \begin{tabular}{|c|c|c|c|c|c|}
\hline
form & $a_+$ & $\lambda_+$ & $c_+$
 & $10^5~R_{+1}^{K\pi}$  & $10^3~R_{+2}^{K\pi}$\\
     &       &             &  [GeV$^{-4}$] &  [GeV$^{-2}$] & \\
\hline
Eq.~(\ref{fitfpsimple}) &$\equiv 1$      & $0.0245\pm0.0006$   & $\equiv
0$&&\\
Eq.~(\ref{fitfpsimple}) &$1.000\pm0.004$ & $0.0245\pm0.0015$   & $\equiv
0$&&\\
Eq.~(\ref{fitfpsimple}) &$\equiv 1$      & $0.0238\pm0.0017$
&$0.5\pm1.2$&&\\
Eq.~(\ref{fitfpsimple}) &$1.008\pm0.009$ & $0.0181\pm0.0068$
&$2.8\pm2.8$&&\\
\hline
Eq.~(\ref{fitCPLEARchpt}) & $1.008\pm0.008 $ &
$0.0180\pm0.0067$&$2.7\pm3.0$ &
   $-4.3\pm2.5$ & $0.19\pm0.21$\\
Eq.~(\ref{fitCPLEARchpt}) & $\equiv 1$ & $0.0236\pm0.0019$  & $0.4\pm1.2$ &
   $-2.2\pm0.7$ & $0.02\pm0.09$\\
Eq.~(\ref{fitCPLEARchpt}) & $\equiv1$  & $0.0201\pm0.0006$  &$3.2$ &
   $-4.7\pm0.5$ & $\equiv0.22$\\
Eq.~(\ref{fitCPLEARchpt}) & $1.009\pm0.004$ & $0.0170\pm0.0015$ &$3.2$ &
   $-4.7\pm0.5$ & $\equiv0.22$\\
\hline
  \end{tabular}
  \caption{The fits to the CPLEAR data of various theoretical forms of
$f_+$.
In the last four fits, which use th ChPT results,
 $\lambda_+$ and $c_+$ are derived quantities.
The symbol $\equiv$ means this quantity was set to this value in the fit.}
  \label{tab:fitsfp}
\end{table}
Notice that the fits that go beyond the linear approximation and leave the
normalization free give a significantly lower $\lambda_+$ and with larger
errors. It is within its errors compatible with the linear fit, but
outside the errors from the linear fit, we consider that result from the
fit with curvature to be more reliable. The shift is of similar
size to that observed in Table 1 in \cite{Gjesdal76}.

\subsection{\boldmath $K^+_{e3}$}

There are recent high statistics experiments that show  plots of $f_+(t)$.
They are \cite{Ajinenko02,Levchenko02}.  We will dicuss the data from
\cite{Levchenko02} below.

\subsection{$\lambda_0$ and $\lambda_+$}

Most experiments have analyzed their data assuming linear form-factors.
In Table~\ref{tab:lam+0} we have quoted the PDG2002 values and the
more recent experiments not included in it. We will not use these
numbers much, given the possible shifts when introducing a curvature
in the analysis.
\begin{table}
  \centering
  \begin{tabular}{|c|c|c|c|}
\hline
Process & Ref. & $\lambda_+$ & $\lambda_0$ \\
\hline
$K^+_{\mu3}$ & \cite{PDG2002}        & $0.033\pm0.010$ & $0.004\pm0.009$\\
$K^+_{\ell3}$ & \cite{PDG2002}$\mu e$& $0.0282\pm0.0027$ & $0.013\pm0.005$\\
$K^0_{\mu3}$  & \cite{PDG2002}       & $0.033\pm0.005$   & $0.027\pm0.006$\\
$K^0_{\ell3}$ & \cite{PDG2002}$\mu e$& $0.0300\pm0.0020$ & $0.030\pm0.005$\\
$K^0_{e3}$   & \cite{PDG2002}        & $0.0291\pm0.0018$ & -- \\
$K^+_{e3}$   & \cite{PDG2002}        & $0.0278\pm0.0019$ & -- \\
$K^+_{e3}$   & \cite{Ajinenko02}     & $0.0293\pm0.0025$ & -- \\
$K^+_{\mu3}$  & \cite{Ajinenko03}    & $0.0321\pm0.0045$ &
$0.0209\pm0.0045$ \\
$K^+_{e3}$    & \cite{Levchenko02}   & $0.0278\pm0.0023$ & -- \\
$K^0_{e3}$   & \cite{Tesarek99}      &$0.02748\pm0.00084$& -- \\
\hline
  \end{tabular}
  \caption{The PDG averages for $\lambda_+$ and $\lambda_0$ and the values
from
the most recent experiments. $\mu e$ means that lepton
universality has been used in the measurement.
The result of \cite{Levchenko02} is an update of \cite{Shimizu00} which was
included in the PDG averages. The last result \cite{Tesarek99} is
preliminary.}
  \label{tab:lam+0}
\end{table}

\section{Inputs}
\label{inputs}
\setcounter{equation}{0}

As relevant combinations we have obtained in our earlier work \cite{BT1}
experimental values for $R_{V2}^\pi$ leading to
\be
R_{+2}^{K\pi} = (0.22\pm0.02)~\cdot 10^{-3}
\ee
and
\be
L_9^r = (5.93\pm0.43)~\cdot 10^{-3}
\ee
which used the estimate $R_{V1}^\pi = -0.49~\cdot 10^{-5}~\mbox{GeV}^2$.
The two resonance estimates for the $R_{V1}^M$ quantities done in
\cite{BT1,BCT} via naive vector-meson dominance and the chiral inspired
variety lead to
basically the same estimate
\be
\label{estimateR+2}
R_{+1}^{K\pi} \approx -4~\cdot 10^{-5}~\mbox{GeV}^2\,.
\ee

For inputs for the other parameters we use our fits that include the latest
$K_{\ell4}$ data \cite{Pislak}. These are the $p^4$ fit, and fit 10 to 13
in Table 2 in \cite{ABT3}. This is a reasonable variation of the various
input parameters.

We use the PDG2002 mass values for all the particles involved and
\be
F_\pi = 92.4~\mbox{MeV}\,,\quad F_K/F_\pi =1.22\,.
\ee
The amplitudes for the decays $K^+_{\ell3}$ are calculated with the
mass of the charged kaon and the neutral pion.
Those for $K^0_{\ell3}$ with the mass of the neutral kaon and the charged
pion.

The scale we use in all the coupling constants and the loop integrals
is $\mu=m_\rho=700$~MeV. Almost all conclusions are done from
experimental determinations of the various parameters so the
results are $\mu$-independent.

\section{Numerical Results}
\label{numerics}
\setcounter{equation}{0}

\subsection{Size of the pure loop contributions}

\begin{figure}
  \centering
  \includegraphics[angle=270,width=0.8\textwidth]{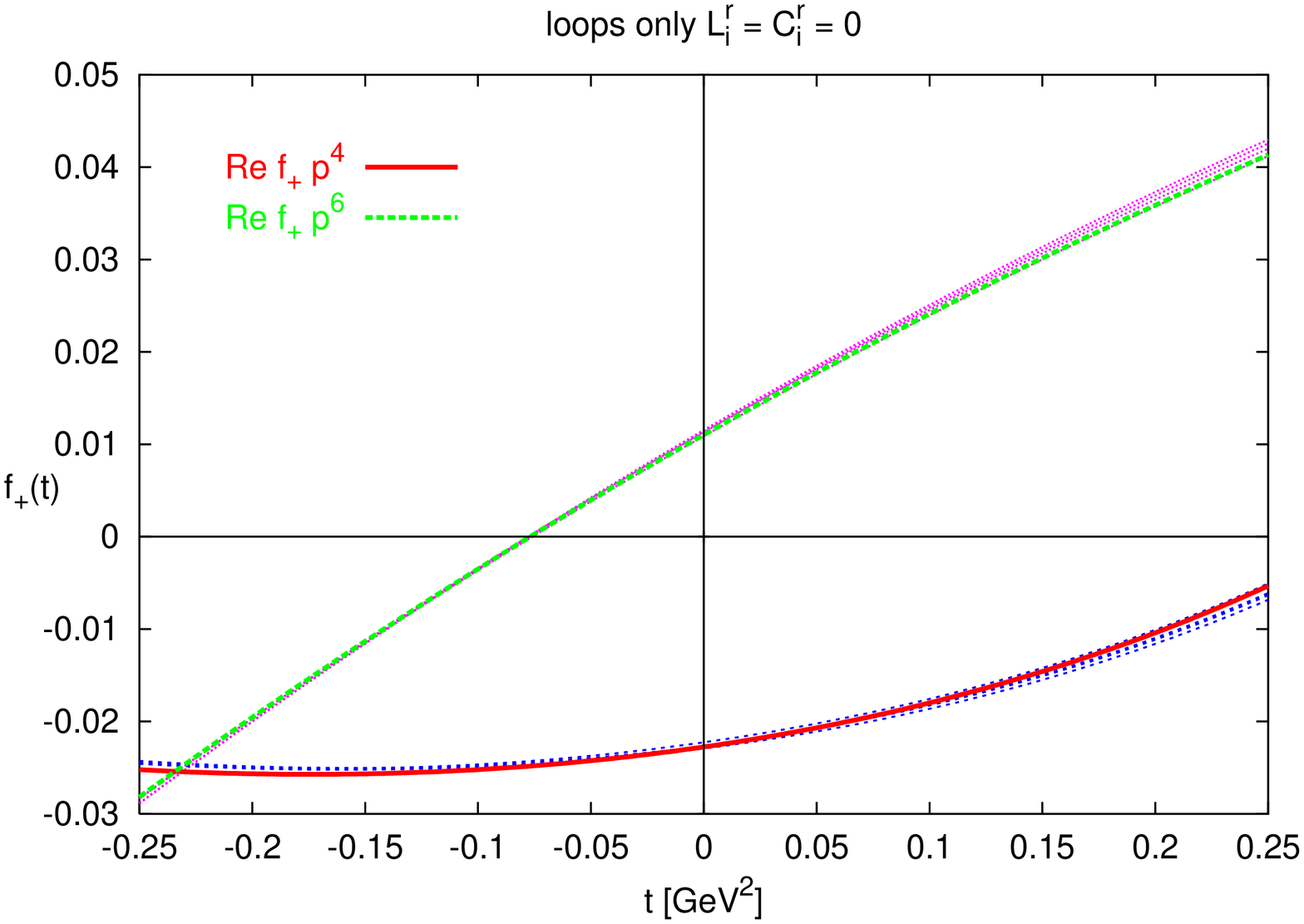}
  \caption{The pure loop diagram contributions to $f_+(t)$. Shown are
  the $p^4$ and the $p^6$ results for several different choices of the pion,
kaon and eta masses.}
  \label{fig:licizerofp}
\end{figure}

\begin{figure}
  \centering
  \includegraphics[angle=270,width=0.8\textwidth]{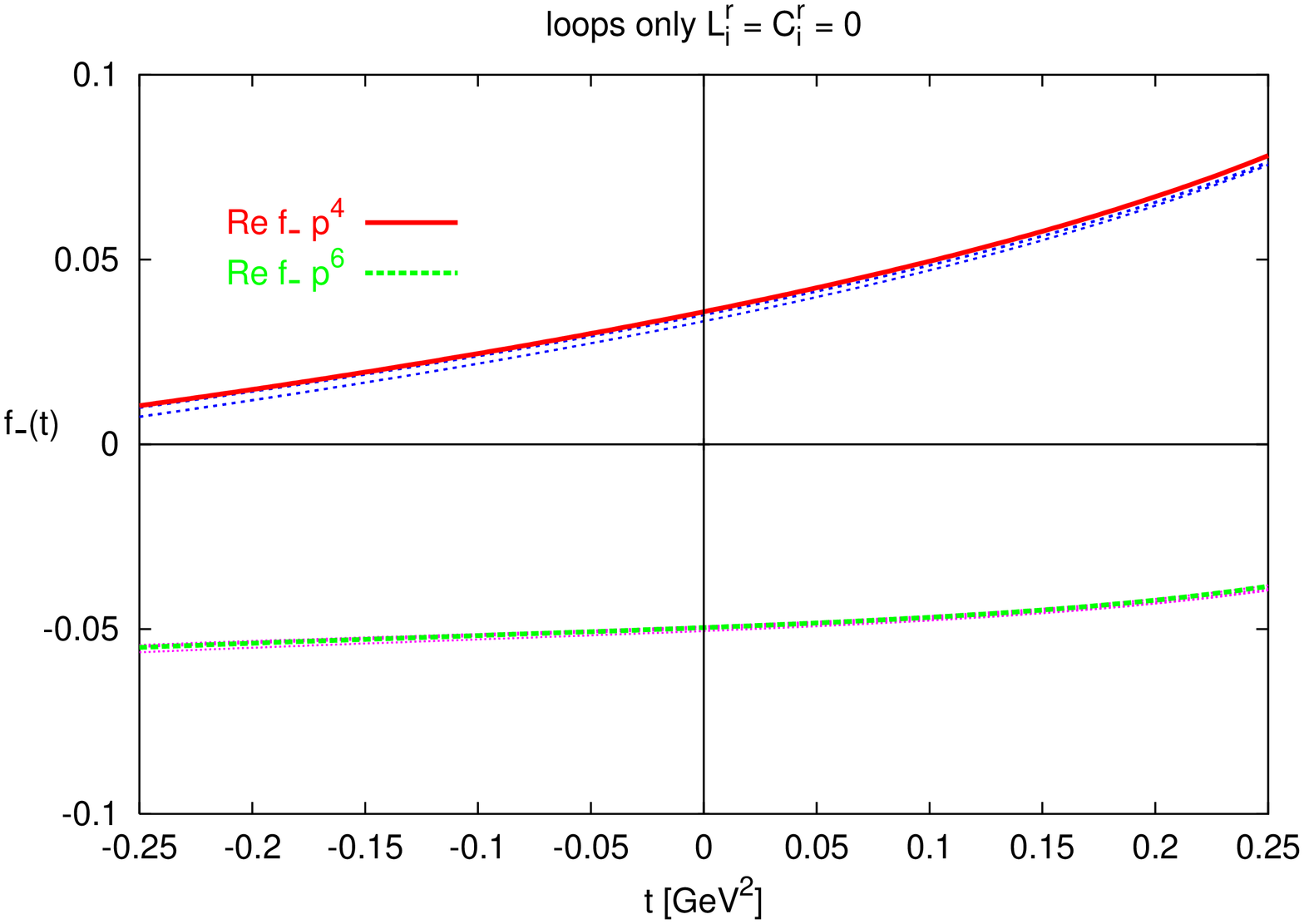}
  \caption{The pure loop diagram contributions to $f_-(t)$. Shown are
  the $p^4$ and the $p^6$ results for several different choices of the pion,
kaon and eta masses.}
  \label{fig:licizerofm}
\end{figure}

In Fig.~\ref{fig:licizerofp} we show the results from the pure loop diagrams
for $f_+(t)$. The different lines are for different choices of
pion, kaon and eta masses. They give some indication of the size of
quark-mass isospin breaking, but it does not include the enhanced
effect discussed in \cite{LR,GL2}.
In Fig.~\ref{fig:licizerofm} we show the equivalent results for $f_-(t)$.

\subsection{Comparison with Refs.~\cite{Post3,Post2}}

At this point we should also compare with the calculation of \cite{Post3}.
In that paper numerical results are quoted in Eqs.~(91-94).
We agree, {\em if we use input $L_i^r$ and masses the same as theirs},
well with their numerical expressions for
$\Delta^{\mbox{\tiny loop}}_{p^6}f_-$ (their Eqs. (92) and (94))
and with their numerical value for $\Delta^{\mbox{\tiny loop}}_{p^6}f_+(0)$
(their Eq.~(93)). We do not agree with their expression for
the $t$-dependence of $\Delta^{\mbox{\tiny loop}}_{p^6}f_+$, our numerical
values differ by roughly a factor of two. Given the good agreement with the
other values this is rather puzzling.

We have attempted to trace
the possible sources of discrepancy, and the present conclusions are:
{\it i)}
recalculating with all $L_i^r =0$
or only  $L_9^r=0$ does not lead to agreement.
If we use as input instead $L_9^r=0.0082$ we get good agreement
with $f_+(t)$, but it spoils the agreement for $f_-(t)$.
{\it ii)} We have used a different split for the reducible and irreducible
parts of the integrals and a different subtraction scheme. Comparing direct
subresults is therefore rather difficult. We have also compared
our results for the pion electromagnetic form-factor \cite{BT1} with those
in \cite{Post2}. We have a small discrepancy for the real part there but a
rather large discrepancy in the imaginary part.
The imaginary part as calculated in \cite{BT1} would give a phase
of a few degrees in \cite{BT1} while those of \cite{Post2} give a phase
significantly above 10 degrees towards 500 MeV. The latter is not
compatible with the expected perturbative buildup of the phase from ChPT.
{\it iii)} The main expected source of discrepancy
is the
slightly different renormalization scheme. The quantity we disagree most
strongly on has a strong cancellation for the value of the $L_i^r$ used
between the pure loop part and the $L_i^r$ dependent part, possibly
amplifying
differences in the result.

\subsection{$f_+(t)$ and Comparison with the CPLEAR and KEP-PS E246  data}
\label{valuel+}

The numerical expression for the $p^6$ contribution with the $C_i^r=0$
and the $L_i^r$ from fit 10 is
\ba
\lefteqn{\frac{1}{F_\pi^4}\left(f_i^L(t)+f_i^B(t)+f_i^H(t)+f_i^V(t)\right)
  }&&\nonumber\\
&=&0.01462+  0.0896353 t +  0.0006313 t^2 + 0.3414 t^3 \quad [K^0_{\ell3}]
\nonumber\\
&=&0.01424+  0.0840569 t +  0.0071463 t^2 + 0.3493 t^3 \quad [K^+_{\ell3}]
\ea
with $t$ expressed in GeV$^2$ and it is valid in the range
$0\le t \le 0.13$~GeV$^2$.

We now compare our ChPT expression at order $p^6$ to the CPLEAR data
\cite{Apostolakis00}. The latter data are normalized to one assuming
a linear dependence. It is therefore that the polynomial fits
done in Sect.~\ref{dataK0e3} added a normalization factor as well.
We now perform a fit using the inputs for the $L_i^r$ from fit 10.
The other choices of the $L_i^r$ give essentially similar results.
So we fit the CPLEAR data to
\be
\label{fitCPLEARchpt}
f_+(t) = \frac{a_+}{\Delta(0)}\left(1+f_+^{(4)}(t)+f_+^{(6)}(t)\right)\,.
\ee
The effect of $R_{+0}^{K\pi}$ of Eq.~(\ref{fpCi}) goes into $a_+$
while $R_{+2}^{K\pi}$ gives the $C_i^r$ part of $c_+$ and
 $R_{+1}^{K\pi}$ gives the $C_i^r$ part of $\lambda_+$ ($c_+$ and
$\lambda_+$
are defined in Eq.~(\ref{fitfpsimple})).

Notice that the fitted value, using the input from the pion electromagnetic
form-factor for $R_{+2}^{K\pi}$,
 gives a value for $R_{+1}^{K\pi}$ in good agreement with the
naive expectation. Notice also that the presence of a curvature
does change the fitted value of the normalization by a little less than
one \%.
A rather important change due to the inclusion of the curvature is the
effect on the value of the slope.  Notice that, using the ChPT expression
and the curvature as determined from the electromagnetic form-factor
leads to
\be
\lambda_+ = 0.0170\pm0.0015\,.
\ee
This value comes from the ChPT in the following way
\be
\lambda_+ = 0.0283 (p^4) + 0.0011 (\mbox{loops }p^6) -0.0124 (C_i^r)\,.
\ee
The $p^6$ correction is about 30\%. The difference with the conclusions
on $\lambda_+$ of \cite{Post3} is to a large extent due to
their fixing the normalization at one.

\begin{figure}
  \centering
  \includegraphics[angle=270,width=0.8\textwidth]{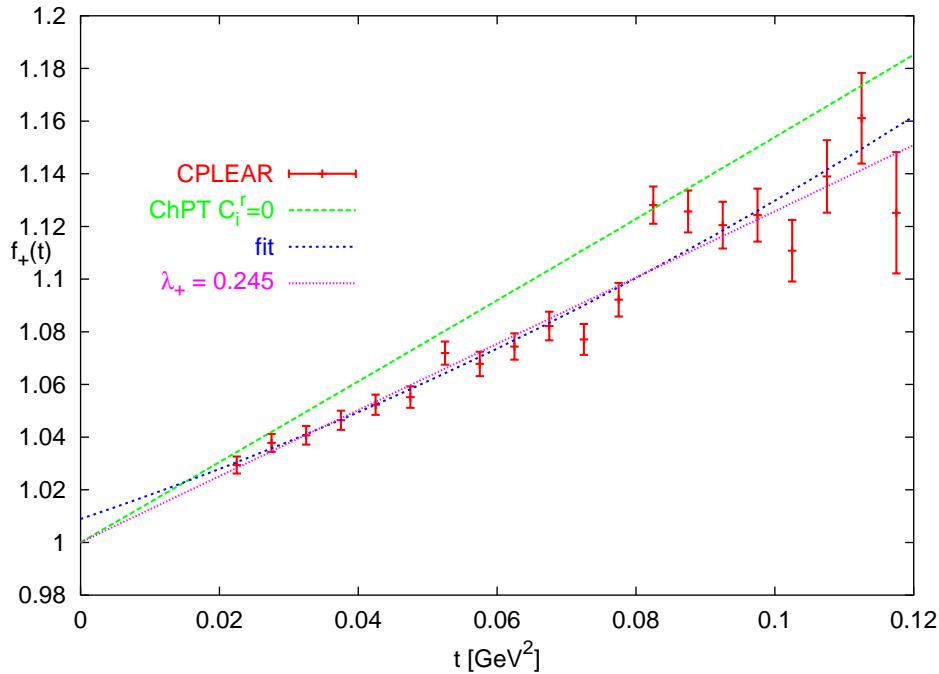}
  \caption{The CPLEAR data together with the normalized ChPT result
without the $C_i^r$,
the last fit reported in Table~\ref{tab:fitsfp} and the linear fit done
by the CPLEAR collaboration.}
  \label{fig:fitsCPLEAR}
\end{figure}

The CPLEAR data together with the normalized ChPT result without the $C_i^r$
contribution,
the last fit reported in Table~\ref{tab:fitsfp} and the linear fit done
by the CPLEAR collaboration, is shown in Fig.~\ref{fig:fitsCPLEAR}.

\begin{figure}
  \centering
  \includegraphics[angle=270,width=0.8\textwidth]{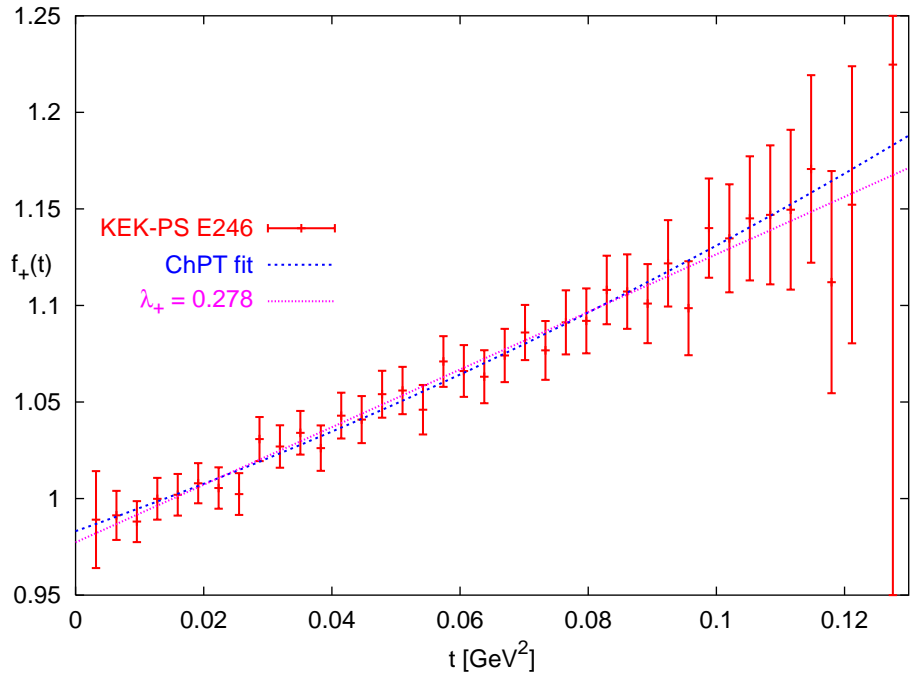}
  \caption{The KEK-PS E246 data together with the ChPT result,
with $R_{+1}^{K\pi}$ fitted and the linear fit of the KEK-PS collaboration.}
  \label{fig:PSE246fit}
\end{figure}

We have performed a similar exercise for the $K^+_{e3}$ data from the
KEP-PS E246 experiment. This is shown in Fig.~\ref{fig:PSE246fit}.
Again, both the linear fit and the one with the curvature fixed
from the pion electromagnetic form-factor
have a similar $\chi^2$. The difference in normalization, relevant for
$|V_{us}|$ is about 0.6\% and
\be
R_{+1}^{K\pi} = -2.5\cdot 10^{-5}\,.
\ee
The value for $R_{+1}^{K\pi}$ is somewhat different from the
value determined from the CPLEAR data, but still compatible with the
resonance
estimate. The value for the slope is
\be
\lambda_+ = 0.0214\pm0.0018\,.
\ee
Notice that in both experiments, KEP-PS E-246 and CPLEAR,
we have neglected the experimental
systematic errors. A possible discrepancy can only be put in after
a full experimental analysis is performed. But again, both the normalization
and slope are changed significantly from the linear fit case.

\subsection{The scalar form-factor $f_0(t)$}

The scalar form-factor as we have shown above is important
since it can be used to determine the $p^6$ constants needed to
evaluate $f_+(0)$.
In this section we show numerical results for $f_0(t)$ and
$\overline\Delta(t)$. We have used here the value
of $F_K/F_\pi=1.22$.

\begin{figure}
  \centering
  \includegraphics[angle=270,width=0.8\textwidth]{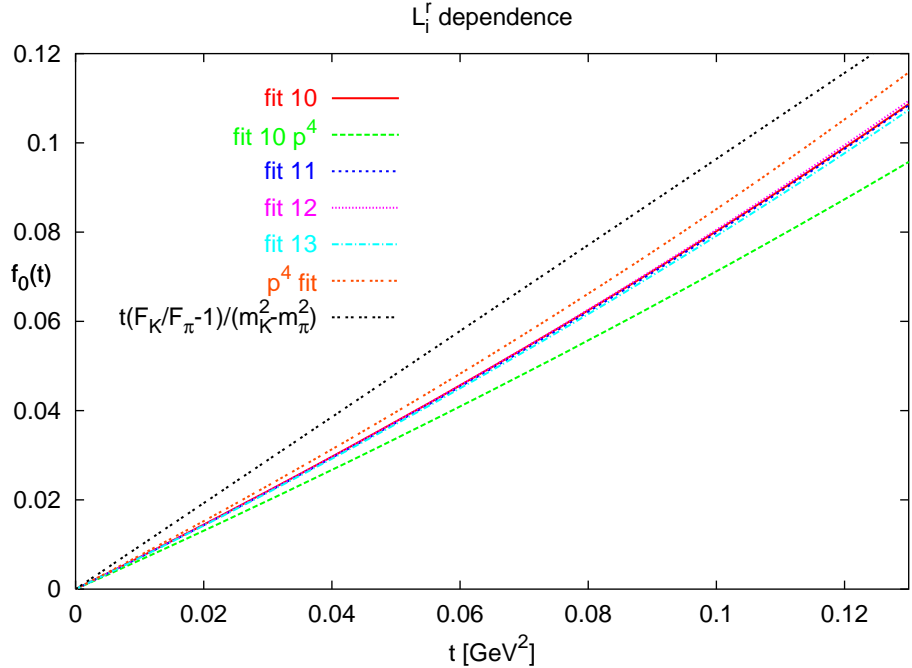}
  \caption{The form-factor $f_0(t)-f_0(0)$
with the $C_i^r=0$. Shown are the cases for the neutral kaon decays
for various sets of the $L_i^r$ together with old current algebra
result.}
  \label{fig:f0t}
\end{figure}

In Fig.~\ref{fig:f0t} we show the function $f_0(t)-f_0(0)$
at order $p^4$ and for the various sets of the $L_i^r$ of
Ref.~\cite{ABT3}. The value of $f_0(0)=f_+(0)$ is discussed in
the next subsection.

As can be seen the convergence from $p^4$ to $p^6$ is quite good.
These are the curves labeled ``fit 10 $p^4$'' and ``fit 10''.
Notice that all fits of the $L_i^r$ done at order $p^6$ (fit 10-13) give
basically identical results. The fit of the $L_i^r$ at $p^4$ (labeled
``$p^4$
fit'')
deviates somewhat but this we consider an artefact as discussed above.
For comparison we have shown the part due to
$t(1-F_K/F_\pi)/(m_K^2-m_\pi^2)$.

A good fit over the entire phase space $0\le t\le 0.13$ ($t$ in GeV$^2$)
is given by
\ba
\overline\Delta(t) &=& -0.25763 t  + 0.833045 t^2 +  1.25252 t^3
\quad [K^0_{e3}]
\nonumber\\
\overline\Delta(t) &=&  -0.260444 t  + 0.846124 t^2 + 1.33025 t^3
\quad [K^+_{e3}]
\ea
The error from the values of the different sets of $L_i^r$
is about 0.0013 at $t=0.13$~GeV$^2$.

We have not attempted to do a fit to any of the data given the
experimental situation on $\lambda_0$. We would however like to point
out that the predicted curvature in $f_0(t)$ is small but of the same order
as in $f_+(t)$. As we saw above, this curvature made a rather large
change in the measured value of $\lambda_+$. A similar effect in
$\lambda_0$ can thus not be excluded and should be studied experimentally.

\subsection{The value of \boldmath$f_+(0)$ and \boldmath$\Delta(0)$}
\label{valuefp0}

The results for $f_+(0)$ with $C_i^r=0$ (which is equivalent to $\Delta(0)$)
are shown in Table~\ref{tab:fp0}. The isospin breaking shown is only an
estimate, we have calculated the $K^+_{\ell3}$ case with the masses
 $m_{K^+}$ and $m_{\pi^0}$ and $K^0_{\ell3}$ with $m_{K^0}$ and $m_{\pi^+}$.
Further work on including isospin breaking fully to two-loop order is in
progress \cite{iso}.
\begin{table}
  \centering
  \begin{tabular}{|c|c|c|}
\hline
 & $K^0_{\ell3}$   & $K^+_{\ell3}$ \\
\hline
$p^4$              & $-0.02266$  & $-0.02275$ \\
$p^6$ loops only   & 0.01130     & 0.01104    \\
$p^6$-$L_i$ fit 10 & 0.00332     & 0.00320    \\
$p^6$-$L_i$ fit 11 & 0.00375     & 0.00355    \\
$p^6$-$L_i$ fit 12 & 0.00216     & 0.00189    \\
$p^6$-$L_i$ fit 13 & 0.00539     & 0.00526    \\
$p^6$-$L_i$ $p^4$ fit & 0.00891  & 0.00863    \\
\hline
  \end{tabular}
  \caption{The various contributions to
$\Delta(0)=\left.f_+(0)\right|_{C_i^r=0}-1.$}
  \label{tab:fp0}
\end{table}
As discussed above, we consider the results with the $L_i^r$ determined
at $p^4$ order rather extreme. We have also investigated how $\Delta(0)$
varies if we vary the $L_i^r$ according to the errors and correlations
determined in \cite{ABT3}. For fit 10, the 68\% CL error gives 0.00124
and for fit 11 it gives 0.00273. Notice that this latter set allows
for a very large variation of $L_4^r$.
We take the latter 0.00273 as a sign of the variation with the $L_i^r$,
notice that includes all the $p^6$ fits given above.
As a conservative estimate of this error we take half of the $p^6$
loop contribution as error and add to it the error from the
$L_i^r$.
We thus obtain
\be
\Delta(0) = -0.0080\pm0.0057[\mbox{loops}]\pm0.0028[L_i^r]\,.
\ee

The value of $f_+(0)$ is related to $\Delta(0)$ via
\be
f_+(0) = 1+\Delta(0)-\frac{8}{F_\pi^4}\left(C_{12}^r+C_{34}^r\right)
\left(m_K^2-m_\pi^2\right)^2\,.
\ee
A naive estimate of $C_{12}^r$ can be made from scalar dominance
of the pion scalar form-factor (SMD) leading to
\be
\left.C_{12}^r\right|_{SMD} = -\frac{F_\pi^4}{8 m_S^4}
\approx -1.0~\cdot 10^{-5}\,.
\ee
The other combination can in principle be estimated from $\lambda_0$
via
\be
\lambda_0 =
\frac{m_\pi^2}{m_K^2-m_\pi^2}\left(\frac{F_K}{F_\pi}-1\right)
+\frac{8m_\pi^2}{F_\pi^4}\left(2C_{12}^r+C_{34}^r\right)
\left(m_K^2+m_\pi^2\right)+m_\pi^2\frac{d}{dt}\overline\Delta(t)\,.
\ee
As an example take
$\lambda_0 = 0.009\pm0.010$ leading to
\be
2C_{12}^r+C_{34}^r = (-1.0\pm1.7)~\cdot 10^{-5}\,.
\ee
Putting both estimates together gives
\be
\left.f_+(0)\right|_{C_i^r} \approx 0.0\pm0.1.
\ee
Essentially a $1\%$ precision on $f_+(0)$ requires a  measurement
of $\lambda_0$ to 0.001 (about 5\%), assuming we can determine $C_{12}^r$
with the relevant precision from other sources \cite{BD}.

The estimate of the $p^6$ corrections given in \cite{LR} is for
the analytic contribution (proportional to $(m_s-\hat m)^2$) and
contributes to the $C_i^r$ dependent part
\be
\left.f_+(0)\right|_{C_i^r} \approx -0.016\pm0.008.
\ee
If we use this value we get for $f_+(0)$ the present best estimate
\ba
\label{fp0final}
f_+(0) &=& 1-0.02266~[p^4]+0.01130~[p^6 ~\mbox{pure loops}]
+0.00332~[p^6 ~L_i^r]-0.016 ~[p^6~\cite{LR}]
\nonumber\\&&
\pm0.0057 [\mbox{loops}]\pm 0.0028~[L_i^r]\pm0.0080~[C_i^r \cite{LR}]
\nonumber\\&&
\hspace{-0.7cm} = \hspace{0.3cm} 0.9760 \pm 0.0102\,,
\ea
where errors have been added in quadrature.
This satisfies the bound $f_+(0) \le 1$ \cite{Furlan} (as quoted in
\cite{LR}).
Notice that this bound when combined with the other results here gives a
constraint on the values of $C_{12}^r+C_{34}^r$ and thus on a
combination of the slope and curvature in the scalar form-factor.
The total value and error on $f_+(0)$ awaits an {\em experimental}
determination of the needed constants but Eq. (\ref{fp0final}) is our
present
best estimate.

The net order $p^6$ contribution
is roughly one order of magnitude smaller, and with the same sign,
as the $p^4$ one, due to the sizable cancellation between the calculation
of the loop contributions and the {\em estimate} of the contribution from
the $p^6$ constants.
We have reevaluated
$\vert V_{us}\vert$ following the basic steps presented in Sec. 7.3 of
\cite{Radiative} but using as input the previous expression
(\ref{fp0final}) obtained for the $K^+$ decay case.
The result should only be taken as preliminary since the full
analysis including
the isospin breaking terms at $p^6$ is still missing as well as an analysis
of the effect of the curvature in the form-factor on the experimental value.
It explicitly contains: the e.m. corrections at one-loop order
and the next-to-next-to-leading order correction in the isospin limit.
\begin{equation}
\label{vus}
\vert V_{us}\vert = 0.2175 \pm 0.0029\,,
\end{equation}
this value is indeed compatible with world-average value $\vert
V_{us}\vert =
0.2196 \pm 0.0023$ within errors. Notice that the theoretical
errors are larger due
to the uncertainty in the $C_i^r$ and they are at the same footing as the 
actual experimental ones.  Another recent evaluation of $V_{us}$
along the same lines can be found in Ref.~\cite{Cirigliano}.

\section{Summary and Conclusions}
\label{conclusions}
\setcounter{equation}{0}

We have performed a calculation to two-loop order of $K_{\ell3}$ decays
in the isospin limit. As far as we have been able to check, this calculation
agrees analytically with the earlier on in \cite{Post3}. We agree with
some of the numerical results of that work but not all.

For $K_{\ell3}$ decay measurements we have shown how the value of $f_+(0)$
needed for the determination of $V_{us}$ can be determined from
the slope and curvature of $f_0(t)$ which can be measured in $K_{\mu3}$.
It is possible that additional information from the pion scalar form-factor
near 0 allows the measurement of the slope only to be sufficient \cite{BD}.

We have presented a present best value for $f_+(0)$ based on an estimate
of the $p^6$ constants $C_{12}^r$ and $C_{34}^r$. It is clear that
this can be further improved after the above measurements are performed.
{}From the calculated best value of $f_+(0)$ we can deduce directly
an improved estimated value of the CKM matrix element $\vert V_{us}\vert$.

As can be seen from our comparison with the data for $f_+(t)$ the presence
of curvature can make a sizable impact both on the determination of the
value of the form-factor at zero and on the slope. The large change in the
value $\lambda_+$ we found is entirely due to this effect and was compatible
with estimates of the $C_i^r$ involved.

\section*{Acknowledgements}
We thank Gabriel Amor\'os for participation in the early stages
of this project and the CPLEAR and KEP PS E246
collaborations for providing us with their data.
This work has been funded in part by
the Swedish Research Council, the European Union TMR
network, Contract No. HPRN-CT-2002-00311  (EURIDICE).
FORM 3.0 has been used extensively in these calculations \cite{FORM3}.
We want to thanks to F. Blanc, P. Post and J. Stern for comments on the
manuscript.

\end{document}